\documentclass[aps,prl,twocolumn,showpacs,preprintnumbers,amsmath,amssymb]{revtex4}
\usepackage{dcolumn}
\usepackage{bm}
\usepackage[dvips]{graphicx, graphics,epsfig}

\begin{document}

\title{On the change in Inertial Confinement Fusion Implosions upon using  an \textit{ab initio} multiphase DT equation of state.}

\author{L. Caillabet, B. Canaud\footnote{Corresponding author: benoit.canaud@cea.fr}, G. Salin, S. Mazevet, P. Loubeyre}
\affiliation{CEA, DAM, DIF, F-91297 Arpajon, France}

\date{\today}

\begin{abstract}
 Improving the description of the equation of state (EoS) of deuterium-tritium (DT) has recently
 been shown to change significantly the gain of an Inertial Confinement Fusion (ICF) target (Hu \textit{et al.}, PRL \textbf{104}, 235003 (2010)).
 We use here an advanced multi-phase equation of state (EoS), based on \textit{ab initio} calculations, to perform a full optimization of
 the laser pulse shape with hydrodynamic simulations starting from 19~K in DT ice. The thermonuclear gain is shown to be a robust estimate over possible uncertainties of the EoS. Two different target designs are discussed, for shock ignition and self-ignition. In
 the first case, the areal density and thermonuclear energy can be recovered by slightly increasing the laser energy. In the second case, a lower in-flight adiabat is needed, leading to a significant delay (3ns) in the shock timing of the implosion.
\end{abstract}

\pacs{52.57.Bc, 47.40.Nm, 47.55.Kf, 52.57.Fg}

\maketitle  
The DT EoS, starting from the cryogenic solid and undergoing a wide range of plasma conditions, is a key input to
simulate the ICF implosion of the DT pellet and hence to quantify the fusion energy production.  The current uncertainty in the EoS of
DT, more particularly in the strongly correlated and degenerate
regime, is maintaining uncertainties in hydro-simulations for the
prediction of ICF thermonuclear gains and also for the shock timing
design. Recently, Hu \textit{et al.}\cite{hu10prl} have shown a
significant reduction (30 \%) of the thermonuclear gain by improving
the DT EoS in the strongly coupled and degenerate regime. A direct
implication of this work is the reduction of the safety margin in
current ICF designs that could be deleterious to achieve high gains.
But, no re-optimization of the laser pulse has been performed with
this partial \textit{ab initio} EoS, only valid for $T\ge 1.35$~eV. Consequently, that poses two important
questions: can the decrease in thermonuclear energy be avoided by
re-shaping the laser pulse, and what result would this have on the
tuning of shock timing? The aim of the present work is to address
these two questions. \\
\indent We use here a recently published Multi-Phase DT
EoS\cite{caillabet11prb}, dubbed MP-EoS, that is based on \textit{ab initio} calculations 
in the strongly coupled and degenerate regime ($\rho=0.5-12.5$~g/cc and $T\le 10$~eV), and extended here to
cover the whole thermodynamical path of DT ice ($\rho=0.25$ -
4.10$^{3}$ g/cc and T=0-10~keV). Hydro-simulations
can thus start at the realistic temperature of 19~K for DT ice, that allows to take into account the
effect of the solid-liquid transition and to avoid any effect of pre-heating of the shell. Two different target designs are considered, one for shock
ignition and the other for self-ignition, corresponding to two different
thermodynamic paths for the compression of the DT fuel. It will be
shown below that the energy gain of the target is in fact a robust
estimate, that can be recovered after optimization of the laser
pulse. But the use of a more realistic EoS can lead to large
changes in the shock timing.

\begin{figure}[!h]
\includegraphics[width=5cm] {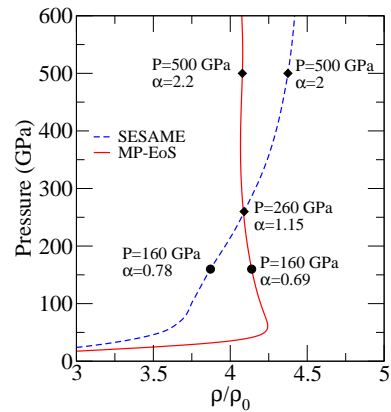}
\caption{Principal Hugoniot curve of the cryogenic DT ($\rho_{0}=0.25$~g/cc, $T_{0}=19$~K) for the MP-EoS and SESAME EoS\cite{KERL03}. The diamonds (resp. the circles) indicate the pressure and the in-flight adiabat behind the first shock induced by the laser pulse in the LMJ target (resp. the Hiper-like target).}
\label{fig1}
\end{figure}

In classical Inertial Confinement Fusion\cite{NUCK72,ATZE04} (ICF),
a single shell of cryogenic DT ice, enclosing the
DT gas, is accelerated inward by laser irradiation in
direct-drive\cite{CANAU07} (DD) or by x-rays in indirect
drive\cite{GIORL06} (ID). A shaped laser pulse creates a multiple
shock train that travels through the target towards the center. To
achieve a net energy gain, a very high-density DT shell should be
assembled by an in-flight compression around the very hot DT gas.
Two quantities are more specifically varied to control the
characteristic of the implosion: the pressure after the first
shock that sets the value of the in-flight adiabat of the compression path
(defined as the ratio of the multiple-shock induced-pressure over
the Fermi pressure : $\alpha=P[GPa]/217\rho^{5/3}[g/cc]$); the
peak implosion velocity of the DT ice shell that
scales with the internal energy given to the DT fuel. During the
implosion, the central gas and the cryogenic DT shell follow two
very different thermodynamic paths. The gas stays in a hot, weakly
coupled and  classical regime (the coupling parameter
$\Gamma=e^{2}/a_{i}k_{B}T \ll 1$, where $a_{i}$ is the ionic sphere
radius, and the degeneracy parameter $\theta=T/T_{F} \gg 1$ with
$T_{F} $ the Fermi temperature of the electron gas). This regime is
well described by physical
models\cite{KERL03,SAUM95,ROSS98,JURA00,CHABR98,POTEK10}. The SESAME
EoS\cite{KERL03}, based on physical models, gives a good description
of the compression of the DT gas during the implosion and is used in
the present work.  On the other hand, the compression of DT ice goes
through strongly coupled and degenerate plasma states, certainly
throughout the deposition time of the laser energy to the pellet
when the multi-shocks cross the target. In this case,
 many-body effects on the EoS  are important and can only accurately be described
by \textit{ab initio} methods; quantum molecular dynamic (QMD) for
low temperatures and Path integral Monte Carlo (PIMC) for high
temperatures. The MP-EoS\cite{caillabet11prb} is at present the
most advanced EoS of DT to treat the states of strong coupling and
degeneracy. The MP-EoS, applied to DT, covers the range 0.5~g/cc -
12~g/cc, and up to $T=10$~eV, and synthesizes the
QMD\cite{LENO00,HOLS07,MORA10} and PIMC
calculations\cite{MILI00,hu10prl} published so far in this domain.
In addition, this EoS takes into account the quantum contributions
of the nuclei. It has been shown to successfully reproduce all of the
experimental data on the EoS of H$_{2}$ and D$_{2}$ published in the solid,
the molecular fluid and the plasma state.  We have extended it by
performing \textit{ab initio} calculations down to $\rho=$~0.25~g/cc in the
same range of temperatures, in order to have an \textit{ab initio}
description of the cryogenic DT at the initial condition of 19~K. The
MP-EoS converges to Chabrier and Potekhin's
model\cite{CHABR98,POTEK10} in the limits of the domain of \textit{ab initio} calculations, and this model is then
used to extend the MP-EoS over the thermodynamical range of the
compression of the DT ice shell, i.e up to densities of $10^{3}$
g/cc and temperatures of 10~keV. The intermediate domain in density between the gas and the
solid is never traversed at low or warm temperature during the
implosion, so an accurate description of this domain is not necessary here.
For a given pressure and temperature, discrepancies in the energy
and in the pressure between the MP-EoS and the SESAME EoS are $\ge$
5\% when $T \le$ 86~eV in the density range 0.25-25~g/cc. That is
illustrated in FIG.\ref{fig1} by comparing the principal Hugoniot
curves derived from the MP-EoS and the SESAME EoS, that shows a
clear difference in compressibility of the DT ice under dynamic
loading. The implication  of such differences in the DT EoS for the
optimization of ICF implosion is discussed below for two ICF
targets, the LMJ baseline direct drive design\cite{canaud04nf} and
a shock-ignited (SI) Hiper-like direct drive
design\cite{canaud10njp}. Hydrodynamic calculations were performed
with the multidimensional Lagrangian radiation-hydrodynamics code
FCI2\cite{fci1} employed here in its one-dimensional version. It
includes inverse-Bremstralhung laser absorption with a one-dimensional
ray-tracing package, flux-limited Spitzer heat conduction,
multi-group radiative transfer, and multi-group fusion-product
transport. Both designs are directly driven by shaped UV-laser pulses.

%------------------
%------------------

%------------------
\begin{figure}[!h]
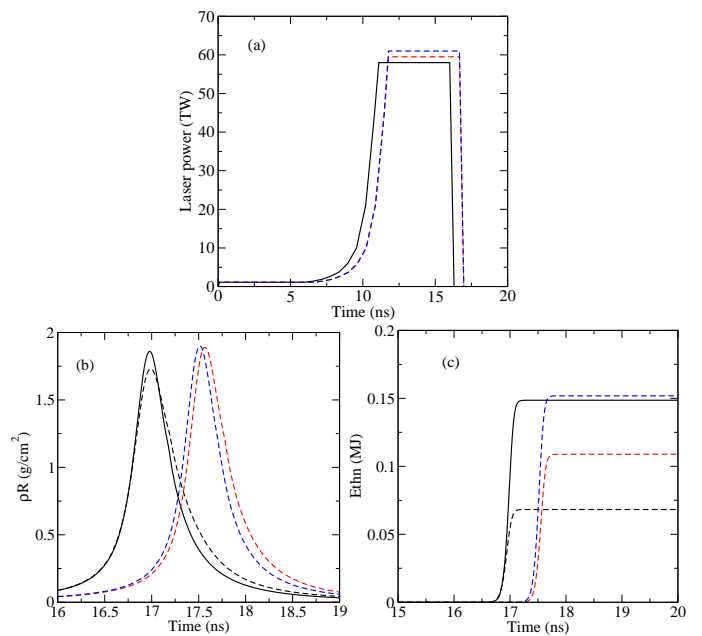

\centerline{\includegraphics[width=4.5cm] {figure2a.eps}}
\centerline{\includegraphics[width=4.5cm] {figure2b.eps}\hfill \includegraphics[width=4.5cm] {figure2c.eps}}
\caption{Output of the hydrodynamic simulations of the HiPER-like target: (a) the laser pulse, (b) the areal density and (c) the thermonuclear energy as a function of time. The solid lines represent hydrodynamic simulations with the SESAME EoS; the dashed lines represent hydrodynamic simulations where the SESAME EoS has been replaced by the MP-EoS, keeping the initial laser pulse (black dashed line), the initial implosion velocity (red dashed line) and the initial thermonuclear energy (blue dashed line).}
\label{fig2}
\end{figure}
%------------------

The Hiper-like target consists of a 316
$\mu$m-thick "all-DT" spherical pellet with a 1244 $\mu$m
inner-radius \cite{canaud10njp}. The results of four
hydro-simulations are presented in FIG.2. First, an optimization of the laser pulse shape is done using the SESAME-EoS, with a
pressure after the first shock of 160 GPa corresponding to an
in-flight adiabat of $\alpha=0.78$, and a peak velocity of 290~km/s. Secondly, the MP-EoS is used keeping the same laser pulse. Thirdly, the laser pulse is redesigned using the MP-EoS and keeping the implosion velocity at 290 km/s. Finally, the drive part of the laser pulse is increased in order to recover the same thermonuclear energy that in the first case but with the MP-EoS. By replacing 
the SESAME EoS by the MP-EoS and keeping the laser pulse optimized with the SESAME EoS, significant differences appear at the beginning of the implosion that act on the in-flight adiabat and  on the
implosion velocity. This results in a large reduction of the peak areal
density at stagnation as well as in the thermonuclear energy (of about
7\% and 60\% respectively). The decrease of the thermonuclear energy qualitatively confirms the less efficient
compression of the DT fuel when strong correlation and degeneracy
effects in the plasma are accurately taken into account, as pointed
out by  Hu \textit{et al.}. That is related to the significant
modification of the thermodynamic path followed by the DT shell
during implosion. In particular, the density and temperature at stagnation are significantly modified (the average density and temperature of the shell at stagnation are of 319~g/cc and 0.63~keV with the SESAME EoS and of 285~g/cc and 0.53~keV with the MP-EoS).

%------------------
\begin{figure}[!h]
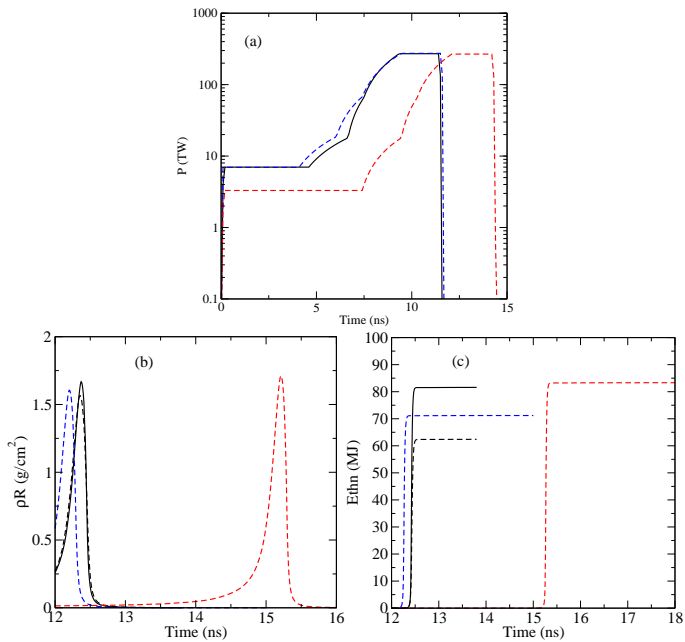

\centerline{\includegraphics[width=4.5cm] {figure3a.eps}}
\centerline{\includegraphics[width=4.5cm] {figure3b.eps}\hfill \includegraphics[width=4.5cm] {figure3c.eps}}
\caption{Output of the hydrodynamic simulations of the LMJ-baseline DD target: (a) the laser pulse, (b) the areal density and (c) the thermonuclear energy as a function of time.  The solid lines represent hydrodynamic simulations optimized with the SESAME EoS at a first shock pressure of 500~GPa; the black dashed lines represent hydrodynamic simulations where the SESAME EoS has been replaced by the MP-EoS, keeping the initial laser pulse; the blue dashed lines represent hydrodynamic simulations optimized with the MP-EoS at a first shock pressure of 500~GPa; the red dashed lines represent hydrodynamic simulations optimized with the MP-EoS at a first shock pressure of 260~GPa.}
\label{fig4}
\end{figure}
%------------------

As the implosion history is connected to shock propagation, an
accurate shock timing analysis is needed. The foot of the laser
pulse creates a first shock at a pressure of 160~GPa which
corresponds to a 6.6\% higher density when using the MP-EoS instead
of the SESAME EoS (see FIG.\ref{fig1}). Consequently, the secondary
shock propagates faster in the already compressed shell of DT and catches
up the first shock created by the foot. This leads to a
shock mistiming which can easily be corrected by delaying
the main drive of the laser pulse in time. Indeed, in order to maximize the
peak areal density, the main peak of the laser pulse has to be
delayed by 670~ps. With this optimized shock timing, a slight
decrease (1\%) in the peak implosion velocity v$_{imp}$ is still
observed. Since the peak kinetic energy is completely transformed into
internal energy during deceleration, this leads to a reduced
thermonuclear energy. The same implosion velocity can be recovered
by slightly increasing the main drive power (from 357 kJ to 364 kJ).
With the revised laser pulse shown in FIG.\ref{fig2}(a), the value
of the areal density is recovered with the SESAME EoS but a
significant reduction of the thermonuclear energy of about 27\% is
still observed (see FIG.\ref{fig2}(b) and (c)). To recover the
thermonuclear energy, the laser energy has to be further increased to
374 kJ . For all these cases, the solid-liquid phase transition appears to have no effect. This is confirmed by simulations started at a higher temperature (30~K) than the triple point temperature which give the same results. In conclusion, using the more realistic MP-EoS, the shock
timing of the laser pulse has to be modified by 670 ps and the
energy of the laser increased by 5\% in order to recover the
performance of the target calculated using the SESAME EoS.

The Direct-Drive baseline design for LMJ target is a massive 1642 $\mu$m outer radius, 1 $\mu$m-thick CH-layer,
202 $\mu$m-thick wetted-foam and 164 $\mu$m-thick DT ice inner-layer
capsule\cite{canaud04nf}. Four hydro-simulations output are performed
and presented in FIG.3. As previously, the first one concerns the optimization done with SESAME-EoS. The
pressure after the first shock is 500 GPa that corresponds to an
in-flight adiabat of $\alpha=2$. The peak implosion velocity
is 400 km/s and a thermonuclear energy of 82~MJ is obtained for a 1D-incident laser
energy of 1 MJ. The second case concerns the implosion done with the same laser pulse shape and the MP-EoS instead of the SESAME EoS. As for both
Hiper-like target discussed above and Hu's work, a reduction ($\sim$ 7 \%) of the areal density
and a significant decrease ($\sim$ 30 \%) of the output thermonuclear energy are observed. The third case concerns the optimization of the laser pulse using the MP-EoS and keeping the peak implosion velocity at 400 km/s. The laser ramp is modified as
shown in FIG.3 and  an optimum is achieved with a thermonuclear energy of 71~MJ. The peak areal density occurs 200 ps earlier than in the first case mainly due to the higher compressibility of the SESAME-EoS compared to the MP-EoS at 500 GPa. Increasing the peak
implosion velocity increases the thermonuclear energy to a maximum of 77~MJ which is still 6\%
below the SESAME result. In fact, the MP-EoS
produces much more entropy along a compression path than the SESAME EoS.
The only way to counterbalance this increase consists in chosing a lower infight
adiabat. This is performed in a fourth case  by tuning the post-first shock pressure to 260 GPa (crossing point of the Hugoniot
curves  of the SESAME and the MP EoS, as shown in FIG.1) and leads to a
lower  in-flight adiabat of  1.15 instead of 2.2. The laser
pulse shape is then optimized to obtain the same implosion velocity of 400
km/s. The calculations produces a 1D-thermonuclear energy of 82~MJ for a laser
pulse energy of 971 kJ, hence recovering the previous value
obtained with the SESAME EoS. But in that case, a strong
difference in the shock timing is observed. The peak areal density
is delayed by 3 ns. In addition, a lower in-flight
adiabat could deteriorate the hydrodynamic stability of the implosion. However,
that is beyond the scope of the present work.

Looking at the deceleration phase illustrates the difference between both EoS and their impact on the conversion of the peak kinetic energy into internal energy of the fuel. When the laser ends, the kinetic energy reaches a maximum and the deceleration phase begins. During this stage, the kinetic energy is converted in internal energy of the hot spot and the shell, and this transfer is strongly coupled to the EoS. The comparison of the implosion velocity versus time from its maximum
value to stagnation in two cases, either when thermonuclear fusion reactions are turned on or off, allows to estimate the kinetic energy of the DT shell in excess when ignition occurs.

%------------------
\begin{figure}[!h]
\includegraphics[width=7cm] {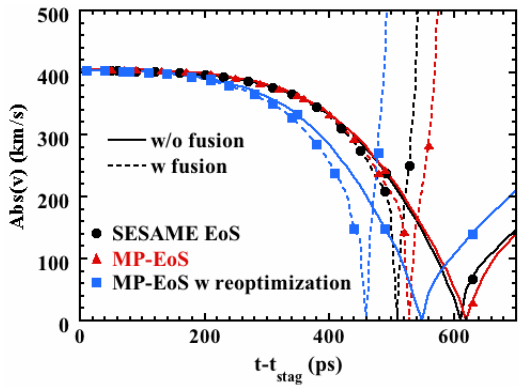}
\caption{Deceleration of the LMJ-baseline DD target with and without nuclear reactions.}
\label{fig5}
\end{figure}
%------------------

Stagnation occurs earlier with thermonuclear reactions than without as shown in FIG.\ref{fig4}. The kinetic energy margin is defined as the ratio of the kinetic energy of the DT shell without fusion at the time of stagnation of the implosion with fusion over the peak kinetic energy.
As seen in FIG.\ref{fig4}, the SESAME case leads to
a kinetic margin of  27 \% and  the MP-EoS case of 21\% before optimization. After reoptimization, the margin grows to 24\%. With the same kinetic energy, The SESAME EoS is more efficient than the MP-EoS to convert kinetic energy into ionic temperature (T$_i$) and areal density
($\rho$R) at stagnation. The ignition thermodynamic conditions (T$_i>$ 6 keV and $\rho$R $>$ 0.2~g/cm$^2$) are met latter with the MP-EoS than with the SESAME EoS. This indicates a greater entropy change during the compression with the MP-EoS.

In conclusion, we have shown the change in the implosion
characteristics for two ICF targets, using an improved treatment for
 the effect of strong coupling and degeneracy in the
equation of state of DT.  A newly developed MP-EoS was used to
perform a full optimization of the ICF implosion under a given laser
energy, starting from the cryogenic conditions of 19~K. For a given
density/temperature condition, the MP-EoS has a higher entropy than
the SESAME EoS commonly used. If the in-flight adiabat is low
enough, slightly more energy than is predicted with the SESAME
EoS will be needed to reach a given thermonuclear yield. However, if
the first shock is too high, the thermonuclear yield cannot be
recovered, even with a large increase in the laser input energy. In
this case, the first shock pressure has to be reduced, recovering the
thermonuclear yield without changing the input energy.
Consequently, the present work demonstrates that the  uncertainty in
the DT EoS will not jeopardize predictions for ICF ignition. The use
of a more advanced EoS such as the MP-EoS is expected to be a valuable way to
scale ICF compression with improved predictions for the shock timing which, as
shown here, may change by up to 3~ns.

%\bibliography{fci_paper}

\end{document}